\documentclass[aps,prl,reprint,superscriptaddress]{revtex4-1}

\usepackage{graphicx}
\graphicspath{{Images/}}
\usepackage{dcolumn}
\usepackage{bm}
\usepackage{float}
\usepackage{color}

\begin{document}

\preprint{APS/123-QED}

\title{Energy barriers for collapsing large-diameter carbon nanotubes}%

\author{R. R. Del Grande}
\email{rdgrande@if.ufrj.br}
\affiliation{Instituto de F\' isica, Universidade Federal do Rio de Janeiro, Caixa Postal 68528, Rio de Janeiro, RJ 21941-972, Brazil}

\author{Alexandre F. Fonseca}
\email{afonseca@ifi.unicamp.br}
\affiliation{Applied Physics Department, State University of Campinas, Campinas, SP, 13083-970, Brazil}

\author{Rodrigo B. Capaz}
\email{capaz@if.ufrj.br}
\affiliation{Instituto de F\' isica, Universidade Federal do Rio de Janeiro, Caixa Postal 68528, Rio de Janeiro, RJ 21941-972, Brazil}


\begin{abstract}
Single-wall carbon nanotubes (SWNTs) are best known in their hollow cylindrical shapes, but  
the ground state of large-diameter tubes actually corresponds to a collapsed dumbbell-like structure, where the opposite sides of the nanotube wall are brought in contact and stabilized by van der Waals attraction. For those tubes, the cylindrical shape is metastable and it is interesting to investigate the energy barrier for jumping from one configuration to another. We calculate the energy barrier for SWNT collapse by considering a transition pathway that consists of an initial local deformation that subsequently propagates itself along the SWNT axis. This leads to finite and physically meaningful energy barriers in the limit of infinite nanotubes. Yet, such barriers are surprisingly large (tens of eV) and therefore virtually unsurmountable, which essentially prevents the thermal collapse of a metastable cylindrical at any reasonable temperatures. Moreover, we show that collapse barriers increase counterintuitively with SWNT diameter. Finally, we demonstrate that, despite such huge barriers, SWNTs may collapse relatively easily under external radial forces and we shed light on recent experimental observations of collapsed and cylindrical SWNTs of various diameters. 

\end{abstract}

\maketitle

\begin{figure}[H]
    \centering
    \includegraphics[width=1.0\linewidth]{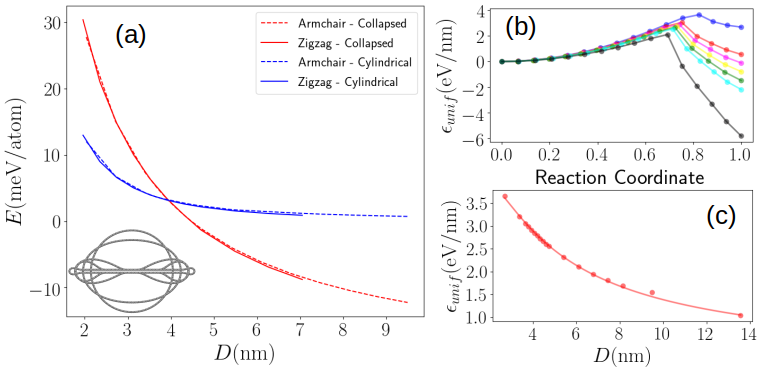}
    \caption{(a) Energy per atom for cylindrical and collapsed SWNTs. The zero energy corresponds to graphene. In the left bottom corner, several snapshots of the uniform collapse of a large-diameter SWNT are illustrated (from circular to oval and then to a dumbbell-like collapsed cross section). (b) Minimum energy pathway obtained by the NEB method for several SWNTs with diameters in the range between 3 and 6 nm. (c) Energy barrier per unit length $\epsilon_{unif}$ as a function of the diameter for the uniform collapse process. The curve is a power-law which gives $\epsilon_{unif}\propto 1/D^{0.76}$.}
    \label{fig:collapso_uniforme_resultados}
\end{figure}

Single-wall carbon nanotubes (SWNTs) are one-dimensional materials that can be thought as graphene sheets rolled up on themselves, typically in a cylindrical shape \citep{SaitoBook}.
However, due to the relatively small flexural rigidity of graphene~\citep{MengJPDAP2013}, it is possible to bring the opposite sides of the nanotube wall into contact and have them stabilized by van der Waals (vdW) interactions. These so-called {\it collapsed} SWNT structures, assume a conformation similar to that of a bilayer graphene with connected edges (Fig. \ref{fig:collapso_uniforme_resultados}(a)) \cite{Zhang2006PRB, Lu2011PRB}.

The relative stability of both structures (cylindrical or collapsed) is governed by the competition between the  attractive vdW interactions and the elastic energy required to bend the nanotube walls. As a consequence, SWNTs with diameters $D \gtrsim 4$ nm are more stable in the collapsed state, with the cylindrical state then becoming metastable in these cases \citep{Zhang2006PRB, Lu2011PRB, HeACSNano2014, HertelPRB1998, TangJournAppPhys2005, TangneyNanoLett2005, HasegawaPRB2006}. Collapsed SWNTs show different electronic, optical and mechanical properties with respect to cylindrical ones, and the reversible transition between cylindrical and collapsed structures is promising to applications as pressure sensors and optomechanical devices \citep{BarzegarNanoLet2016, BarzegarNanoResearch2017, TienchongPRB2008, TienchongNanoLet2010, PerebeinosNanoLett2014, KouAppliedPhysLett2013, BarbozaPRL2008, ParkPRB1999}. It then becomes crucial to evaluate the energy barrier for the transition between the cylindrical and collapsed structures (or vice-versa).

Previous works have attempted to calculate this energy barrier \citep{Zhang2006PRB, Lu2011PRB}. However, they all employed periodic boundary conditions with a minimum-size unit cell along the SWNTs axis. Therefore, it is implicitly assumed that the transition pathway from cylindrical to collapsed tube occurs in a uniform manner along the SWNT, as illustrated in Fig. \ref{fig:collapso_uniforme_resultados}. For short, we call this procedure {\it uniform collapse} of a SWNT. As a result, what is in fact calculated is this case is some sort of ``energy barrier per unit length" and it is easy to see that it will cost an incredibly large amount of energy to collapse a SWNT in this manner if the length of the SWCNT is large (as it typically is). Even worse, in the limit of an infinite SWNT, such energy barrier would be infinite and a SWNT would never collapse.

This suggests that the assumption of a uniform collapse pathway is unphysical. We therefore propose that the collapse of a SWNT will proceed by the nucleation and subsequent growth of an initially local deformation, like a domino process~\citep{TienchongPRB2008,TienchongNanoLet2010}. We call it {\it local collapse}.

In this work, we calculate the energy barrier between the cylindrical and collapsed states of SWNTs as a function of their diameter by considering the local collapse process. Our results reveal an unexpectedly rich and surprising - yet simple - physics: (1) As expected, the local collapse process produces finite barriers even in the limit of infinitely long SWNTs. However, the calculated barriers are huge, making it virtually impossible to thermally collapse a metastable cylindrical SWNT. (2) Energy barriers {\it increase} with SWNT diameter, which is somewhat counterintuitive - as it is well-known that SWNT gets softer radially with increasing diameter - and it is the opposite trend of results from the literature considering uniform collapse~\cite{Zhang2006PRB, Lu2011PRB}. (3) We estimate the critical radial forces required to collapse SWNTs and we show that such forces are relatively small, in apparent contradiction with the large energy barriers. 

We perform both constrained energy minimizations and molecular dynamics (MD) simulations using the LAMMPS package \citep{lammpsPLIMPTON1995} with the AIREBO potential \citep{Brenner2002,Stuart2000} and a cutoff radius of 10.2 $\mathrm{\AA}$ for van der Waals interactions. Most simulations are done for armchair SWNTs, but results are expected to be general since SWNT mechanical properties do not depend substantially on chirality. Details of the calculations can be found the Supplementary Material. 

Fig. \ref{fig:collapso_uniforme_resultados} shows our calculated energy per atom for both cylindrical and collapsed SWNTs as a function of diameter. Both armchair and zigzag SWNT families are shown, and we confirm that the energy per atom is almost independent of SWNT chirality. For cylindrical tubes the energy per atom has a $D^{-2}$ dependence and for collapsed ones the decay goes as $D^{-1}$. Both dependencies are expected from arguments: For the cylindrical case it represents the energy to roll a flat graphene sheet into a cylinder and for the collapsed case it corresponds to the nearly diameter-independent elastic energy excess at the high-curvature regions, which then becomes proportional to $1/D$ after normalization by the number of atoms \citep{TangJournAppPhys2005}. 
The critical diameter $D_c$ for collapse is determined by the crossing of the collapsed and cylindrical curves in Fig. \ref{fig:collapso_uniforme_resultados} and our results indicate $D_c = 3.9$ nm, in consistency with reported theoretical values from the literature \citep{Zhang2006PRB, Lu2011PRB, HeACSNano2014, HertelPRB1998, TangJournAppPhys2005, TangneyNanoLett2005, HasegawaPRB2006}. Interestingly, He {\it et al.} \citep{HeACSNano2014} obtain experimentally that most SWNTs can still survive in the cylindrical shape for diameters near 5.1 nm and, surprisingly, even a few cylindrical  SWNTs as wide as $D\approx 7$ nm can be observed. We address this issue below. 

To confirm the adequacy of the AIREBO potential we evaluate the energy barrier per length for the uniform collapse using the Nudged Elastic Band (NEB) method \citep{NEB1, NEB2, NEB3} (Fig. \ref{fig:collapso_uniforme_resultados}(b)) and we find consistent results with previous works \citep{Zhang2006PRB, Lu2011PRB}: Energy barriers per unit length ($\epsilon_{unif}$) in the few eV/nm range and decreasing with tube diameter (Fig. \ref{fig:collapso_uniforme_resultados}(c)), as SWNTs with larger diameters have smaller radial bending stiffnesses. 

\begin{figure}[H]
    \centering
    \includegraphics[width=1.0\linewidth]{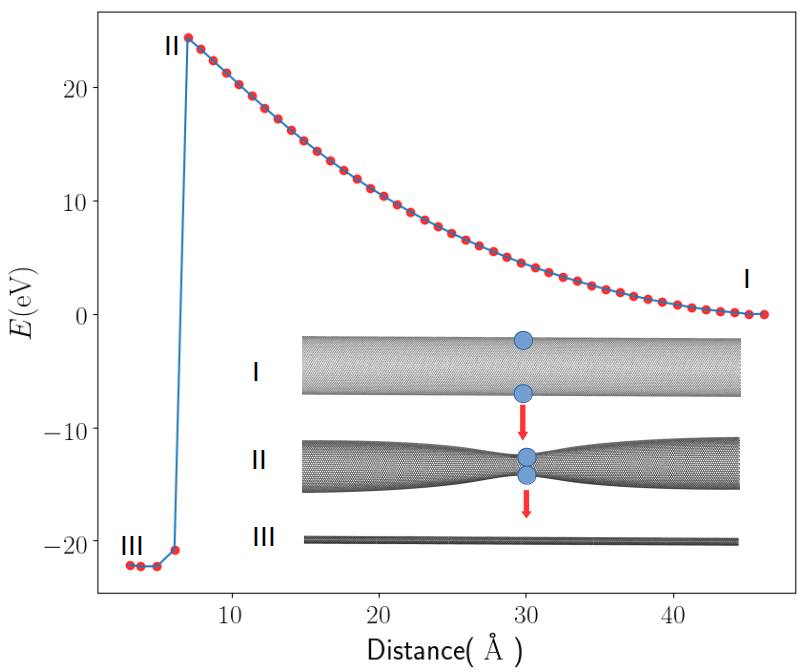}
    \caption{Calculation of the energy barrier using constrained energy minimizations for the local-collapse mechanism. Three snapshots are shown: (I) Initial (cylindrical), (II) saddle-point and (III) final (collapsed). For this particular plot, the SWNT has diameter $D=4.75$ nm and supercell length $L=10$ nm.}
    \label{fig:exemplo_colapso_local}
\end{figure}

Having established that our methodology is robust, we now proceed to calculate the energy barrier for the local collapse mechanism. To induce the local collapse, we construct a SWNT of supercell length $L$ and diameter $D$ using periodic boundary conditions along the tube length. We then perform a series of constrained energy minimizations, as illustrated in Fig. \ref{fig:exemplo_colapso_local}, in which radially-opposite atoms are constrained by a  fictitious spring (with stiffness $10^2$ eV/$\mathrm{\AA}^2$) at decreasing distances, while all other atoms are allowed to relax. A movie showing this process is available in Supplementary Material. The potential energy as a function of distance (Fig. \ref{fig:exemplo_colapso_local}) then increases until the saddle-point is reached (point II in the figure). Beyond that, the potential energy sharply decreases indicating the SWNT collapse. 

\begin{figure}[H]
    \centering
    \includegraphics[width=1.0\linewidth]{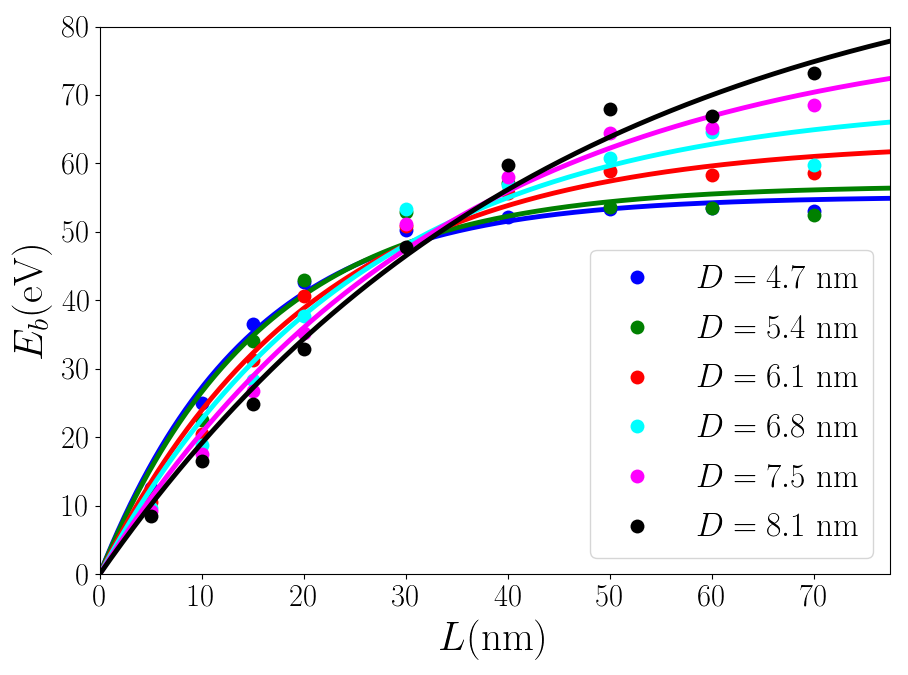}
    \caption{Energy barrier for local collapse as a function of supercell length $L$ for SWNTs of various diameters $D$. The lines are fits to Eq. \ref{expressao_barreia_energetica}.}
    \label{fig:resultados_colapso_local}
\end{figure}

 Using this procedure, we obtain the saddle-point energy, i.e. the true energy barrier within the transition-state theory \citep{TransitionStateTheory}, for the collapse of SWNTs of different diameters $D$ and supercell lengths $L$. Our results are summarized in Fig. \ref{fig:resultados_colapso_local}. Naturally, as SWNTs are typically several $\mu$m in length, we are primarily interested in the energy barrier $E_{\infty}^D$ in the $L \rightarrow \infty$ limit for a given $D$, which is obtained by fitting the various curves in Fig. \ref{fig:resultados_colapso_local} by a saturating exponential function
\begin{equation}
    E_b(D,L) = E_{\infty}^D (1 - \mathrm{exp}(-L/L_C^D)) .
    \label{expressao_barreia_energetica}
\end{equation}

\begin{figure}[H]
    \centering
    \includegraphics[width=1.0\linewidth]{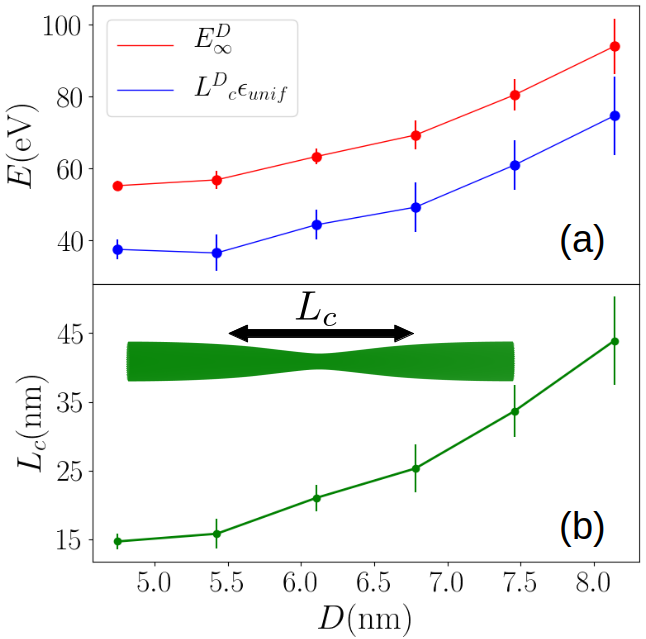}
    \caption{(a) Energy barrier for collapse $E_{\infty}^D$ as a function of diameter in the limit of infinitely long SWNTs (red) and an estimate using the product $L_c\epsilon_{unif}$. (b) Nucleation length $L_c$ of the collapsed region at the saddle-point, as a function of diameter.}
    \label{fig:E_infty_vs_D}
\end{figure}

\noindent The resulting values of $E_{\infty}^D$ are shown in Fig.\ref{fig:E_infty_vs_D}(a). The results are surprising: Energy barriers are in the order of several tens of eV. These are unusually high values for activated processes in condensed matter physics, and certainly much higher than thermal energies for any reasonable temperature, implying that metastable cylindrical large-diameter nanotubes will remain cylindrical essentially forever (as long as they are not mechanically perturbed). Fig. \ref{fig:E_infty_vs_D}(b) shows that $L_c^D$ also increases (superlinearly) with diameter. The length $L_c$ may be interpreted as the typical size of local deformations at the saddle-point or, in other words, the critical nucleus size (or nucleation length) beyond which SWNT collapse will proceed spontaneously. The inset of Fig. \ref{fig:E_infty_vs_D}(b) illustrates the saddle-point deformation for a SWNT with $D=7.5$ nm and $L=70$ nm. The value of $L_c=33$ nm is also indicated in the figure, thus confirming its geometrical interpretation as a nucleation length.

As mentioned, such huge energy barriers are uncommon in condensed matter physics and, as such, they call for a careful confirmation using other methods. In addition, the fact that barriers increase with diameters also requires an explanation, since it is the opposite (and somewhat counterintuitive) trend from the uniform collapse case. A qualitative argument can be constructed by looking precisely at the diameter dependence of $L_c^D$. Even though SWNTs become softer for larger diameters, the nucleation length $L_c^D$ increases faster and compensates the increasing softness. As a matter of fact, a back-of-envelope estimate of the energy barrier for collapse can be made by multiplying the nucleation length $L_c^D$ by the energy per length for uniform collapse $\epsilon_{unif}$, which has a $\approx 1/D^{0.76}$ dependence with diameter (Fig. \ref{fig:collapso_uniforme_resultados}(c)). Since $L_c^D$ has a superlinear increase with $D$, the resulting product also increases with diameter. As shown in Fig. \ref{fig:E_infty_vs_D}(a), this quantity mimics quite well (up to a multiplying factor) the true energy barrier as a function of diameter. 

\begin{figure}[H]
    \centering
    \includegraphics[width=1.0\linewidth]{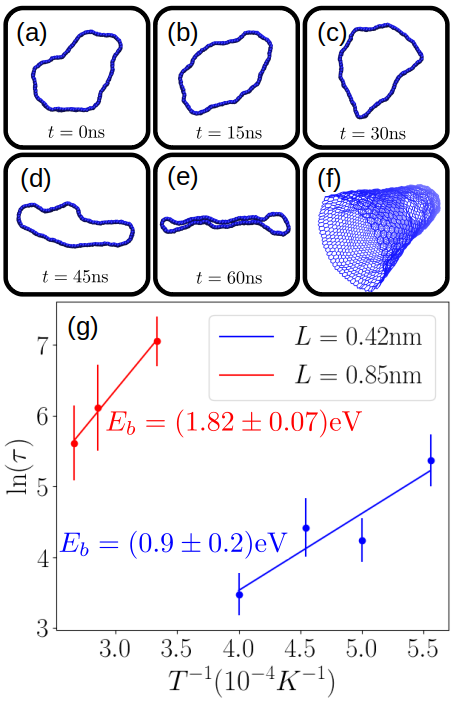}
    \caption{(a)-(e) Snapshots taken at different simulation times (indicated in each panel) for the MD simulation of a SWNT of supercell length $L=0.42$ nm and diameter $D=7.7$ nm in the NVT ensemble at $T=3750$ K. (f) The same SWNT ($D=7.7$ nm), but with a larger supercell length for periodic boundary conditions ($L=10$ nm). (g) Arrhenius plots for the average collapse time of a SWNT with $D = 7.7$ nm and two different values of $L$ SWNT lengths at different temperatures.}
    \label{fig:resultados_MD}
\end{figure}

Another and independent verification of the energy barrier results obtained by constrained energy minimizations can be made by MD simulations. Fig. \ref{fig:resultados_MD} shows snapshots of MD calculations for a SWNT of $D=7.7$ nm with two very different supercell lengths: $L=0.85$ nm (short, Fig. \ref{fig:resultados_MD} (a) to (e)) and $L=10$ nm (long, Fig. \ref{fig:resultados_MD} (f)), both at $T=3750$ K. The short SWNT collapses after a few ns into the simulation, whereas the long SWNT remains cylindrical, even for simulation times into the $\mu$s range. The full videos of these MD simulations are available at the Supplementary Material.

These simulations clearly show, in a qualitative way, the strong dependence of energy barriers on the supercell length $L$, as indicated in Fig. \ref{fig:resultados_colapso_local}. In order to make a more quantitative analysis, we perform MD simulations for several temperatures and two different supercell lengths ($L=0.42$ nm and $L=0.85$ nm). For each condition of $T$ and $L$, an ensemble of ten simulations are performed and the average time for collapse $\tau$ is computed. The results are shown in the Arrhenius plot of Fig. \ref{fig:resultados_MD}(g). The large barriers require long simulation times for the collapse to take place, and temperatures must be kept below 3750 K to prevent sublimation of the SWNTs. Even in these restricting conditions, energy barriers $E_b$ can be obtained from the Arrhenius fit $\tau=\tau_0 e^{E_b/k_BT}$. Our results give $0.9 \pm 0.2$ eV for $L=0.42$ nm and $1.82 \pm 0.07$ eV for $L=0.85$ nm. These are in reasonably good agreement with our results using static minimizations (0.74 eV and 1.48 eV), respectively, thus clearly indicating the validity of both approaches and confirming the $L$-dependence of energy barriers.

From these results, we are confident that our conclusions, although unexpected and surprising, are sound: Collapse barriers of long SWNTs are huge and therefore unsurmountable for any reasonable temperature. As a result, {\it a large-diameter cylindrical SWNT will remain forever cylindrical in a metastable state}. We propose that this is the explanation for the observation of cylindrical SWNTs of very large diameters (some as large as $D\approx 7$ nm) by He {\it et al.}~\cite{HeACSNano2014}. 

\begin{figure}[H]
    \centering
    \includegraphics[width=1.0\linewidth]{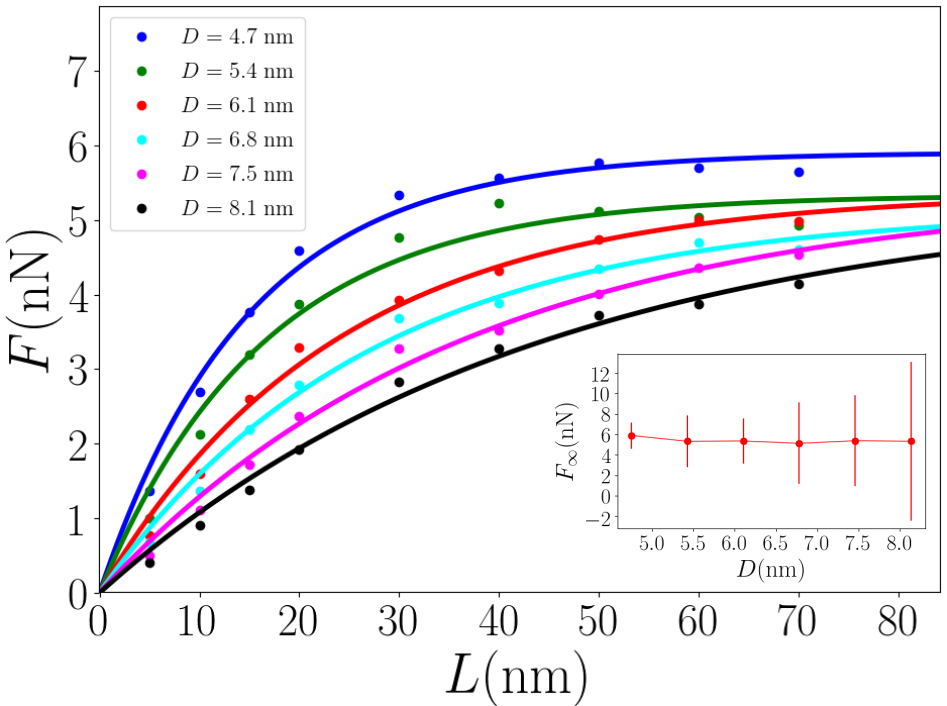}
    \caption{Critical forces for collapse as a function of supercell length $L$ for SWNTs of various diameters $D$. The forces converge exponentially to limiting values $F_{\infty}$ in the limit of large $L$. As shown in the inset, $F_{\infty}$ are roughly independent on diameter.}
    \label{fig:forca_maxima_exercida}
\end{figure}

However, the following question arises: How collapsed SWNTs are formed? 
To help answering this question, we calculate the critical force to induce collapse, simply by taking the derivative of the barrier profiles of Fig. \ref{fig:exemplo_colapso_local}. The results are shown in Fig. \ref{fig:forca_maxima_exercida}. Interestingly, these forces converge exponentially with increasing $L$ to approximately 6 $\mathrm{nN}$, independent of the tube diameter (inset). These are rather ordinary forces in the nanoscale, easily achievable in AFM compression experiments \cite{PalaciPRB2005, BarbozaPRL2009}. Therefore, large-diameter SWNTs can be easily deformed to a collapsed structure by the application an external radial force, even though the energy barriers are huge. These are not contradictory statements: A small force produces large work if applied for a long distance, which is the case for large diameter tubes. This suggests that collapsed SWNTs are likely to be produced by small mechanical stresses that naturally occur during synthesis.

In conclusion, we calculate the energy barriers for collapse of SWNTs by considering a local collapse mechanism. Surprisingly, energy barriers are very large (in the tens of eV range), which essentially prevents thermally-activated collapse of metastable cylindrical SWNTs at any reasonable temperature. Nevertheless, SWNTs can collapse fairly easily by the application of a local radial force. We also demonstrate a counterintuitive increase of energy barriers with nanotube diameter. We find consistent results using both saddle-point determinations by constrained minimizations and MD simulations. Our results reveal a surprising yet simple physics of the mechanical properties of large-diameter carbon nanotubes that can be immediately generalized for any other kind of nanotube. 

We acknowledge the financial support from the Brazilian agencies CNPq, CAPES, FAPERJ and INCT-Nanomateriais de Carbono. AFF particularly acknowledgs grant \#2018/02992-4 from S\~{a}o Paulo Research Foundation (FAPESP). We also thank DIMAT-Inmetro and NACAD/COPPE-UFRJ for the computational resources employed on this work. This research also used the computing resources and assistance of the John David Rogers Computing Center (CCJDR) in the Institute of Physics “Gleb Wataghin”, University of Campinas.

\bibliographystyle{apsrev4-1}

\end{document}